\documentclass[aps,pre,twocolumn,showpacs]{revtex4}
\usepackage{epsfig}
\usepackage{times}
\begin{document}

\title{Selection of noise level in strategy adoption for spatial social dilemmas}

\author{Attila Szolnoki,$^1$ Jeromos Vukov,$^2$ and Gy\"orgy Szab\'o$^1$}
\affiliation{$^1$ Research Institute for Technical Physics and
Materials Science, P.O. Box 49, H-1525 Budapest, Hungary \\
$^2$ ATP-Group, CFTC and Departamento de F\'{\i}sica da Universidade de Lisboa, P-1649-003 Lisboa Codex, Portugal}

\begin{abstract}
We studied spatial Prisoner's Dilemma and Stag Hunt games where both the
strategy distribution and the players' individual noise level
could evolve to reach higher individual payoff. Players are located on the sites of different two-dimensional lattices and gain their payoff from games with their neighbors by choosing unconditional cooperation or defection. The way of strategy adoption can be characterized by a single $K$ (temperature-like) parameter describing how strongly adoptions depend on the payoff-difference. If we start the system from a random strategy
distribution with many different player specific $K$ parameters, 
the simultaneous evolution of strategies and $K$ parameters
drives the system to a final stationary state where only one $K$ value remains. 
In the coexistence phase of cooperator and defector strategies the surviving $K$ parameter is in good agreement with the noise level that ensures the highest cooperation level if uniform $K$ is supposed for all players.
In this paper we give a thorough overview about the properties of this evolutionary process.
\end{abstract}


\maketitle

\section{Introduction}
\label{intro}

Evolutionary game theory has attracted great interest recently
from many scientific fields \cite{maynard_82,hofbauer_98,nowak_06,weibull_95,szabo_pr07}. Physicists, biologists, economists and many other scientists have found it challenging to study multi-agent evolutionary systems.

The Prisoner's Dilemma (PD) game \cite{weibull_95,gintis_00} is an excellent toy model to describe 
the sharpest
conflict situation between individual and common interest as it contains all the basic features of such interactions. The original PD game is a two-person
one-shot game where players can choose between two types of behavior, to cooperate or to defect. They earn payoffs according to the simultaneous decisions of both participants. A cooperator gets the 'sucker's payoff ($S$) against a defector, while successful defection yields the temptation to defect ($T$) for the defector. Mutual cooperation is rewarded by $R$ for each player, while two defectors (receiving payoff $P$) punish each other with the defective behavior. The game is a classical PD game if the payoffs accomplish the relation $T>R>P>S$. This inequality causes two selfish (rational) players to defect independently of the other players decision, thus they get the second worst payoff instead of the second best
one for mutual cooperation resulting in a dilemma situation. 
If $R>T$, the social dilemma is weakened because the unilateral deviation from the mutual cooperation is not beneficial \cite{macy_pnas02}. In the latter so-called Stag Hunt (SH) game to act identically to the partner's strategy would result the highest payoff similarly to the Coordination game.

The introduction of spatiality \cite{nowak_ijbc93} revealed fundamentally new solutions of the game, which cannot be detected in the well-mixed situation. In spatial evolutionary models, players are located on the sites of a network where the links of the network define their possible connections. Players gain their accumulated payoff from games with their immediate neighbors and sometimes - according to the evolutionary dynamics - they can adopt the strategy of a neighbor. Usually the strategy adoption probability depends on the payoff difference and in accordance with the Darwinian principle, individuals with higher payoff (fitness) have a greater chance to supersede the less successful ones. Due to the spatial scenario, cooperation can survive in the system, even if only the simplest strategies are allowed, i.e. unconditional cooperation and defection. Here, cooperators form clusters and support each other through the short range local interactions while defectors punish each other with mutual defection. The invasion processes along the borders of the clusters depend on the irregularity of the interface, the underlying network(s), and the 
evolutionary dynamical rule.

The PD game was studied on many types of networks (different lattices \cite{lindgren_pd94,nakamaru_jtb97,szabo_pre98}, scale-free graphs \cite{santos_prl05}, small world networks \cite{wu_pre05}, etc.), investigating the effect of basic topological features on the measure of cooperation. To reduce the number of degrees of freedom, evolving networks \cite{zimmermann_pre04} were examined, too. In these models, players have the opportunity to change their neighbors during the evolutionary process to attempt to increase their income, i.e., the strategy distribution and the connectivity graph co-evolve. As a result of this evolution, highly cooperative communities could be established.

The simultaneous change of strategy and a player-specific parameter, has been extended to other quantities, too. For example, Fort \cite{fort_epl08} studied models where the elements of payoff matrix were inherited in parallel with the strategy adoption. The evolution of the strategy pass capability \cite{szolnoki_njp08} or the interaction range of players \cite{szolnoki_epl08} are also discussed. Moyano and Sanchez \cite{moyano_jtb09} studied the competition between several pairs of dynamical rules controlling the strategy adoption in the system. The latter results have motivated us to study the evolution of a noise-related parameter \cite{szabo_epl09}, which frequently characterizes the uncertainty in the adoption process \cite{szabo_pre98}. Many things can cause this uncertainty, such as temporal or spatial fluctuations in the payoff values, errors in decision or in perception, emotions, individual point of view (free will), etc.
The role of the noise parameter at different underlying graphs was studied thoroughly in several studies \cite{szabo_pre05,vukov_pre06,vukov_pre08}. It turned out that on structures that can be fully perambulated by stepping only on overlapping triangles (i.e., on structures with triangle percolation), cooperation could be maintained in the widest parameter range when the measure of noise was minimal \cite{szabo_pre05}. While on structures without triangle percolation, the optimal measure of noise for cooperation is shifted to positive values.

In this article, we study the simulatenous evolutions of noise parameter - as the quantity characterizing the adoption rule - and strategy distributions on different, relevant structures. Square lattice and the kagome lattice will be analyzed as good examples for graphs with and without triangle percolation. In our model, players can apply different noise parameter
and during the evolutionary process, they can adopt not only the strategy of another neighboring player but also the individual noise parameter. In other words, a player can learn not only a more successful strategy but also the way how successful player reacts to the
payoff differences, i.e. his adoption rule. The two elementary, evolutionary steps are independent. We will show that as a result of this dynamics, only one noise level
remains in the final state even if several strategies can co-exist in the co-existence region and the remaining 
noise level is close to the one providing the highest cooperation for systems with homogeneous noise distribution. In a previous letter \cite{szabo_epl09} we have considered only the case of weak PD. Now the latter investigation is extended to the parameters over the limits of the weak PD and SH games by revealing the connection of cooperation density surface and the location of fixed noise level.
Finally we briefly discuss what happens if not only the strategies and 
the way of strategy adoption but also the payoff matrix are allowed to adopt throughout the same imitation mechanism. We used Monte Carlo (MC) simulations and an extended version of the dynamical mean-field approximation (detailed here) to perform the investigations.

\section{The model}

In our model, players are located on the sites $x$ of a square (consisting of $L \times L$ sites with periodic boundary conditions) or a kagome lattice ($3\times L\times L$ sites). These interaction graphs can be used as the two representatives of the characteristic features of two-dimensional lattices, i.e. lattices with and without triangle percolation. Players can follow one of the simplest strategies that is unconditional cooperation ($s_x=C$) or unconditional defection ($s_x=D$). They gain their cumulated payoffs from one-shot PD games with their four nearest neighbors. To reduce the necessary parameters we are using a re-scaled payoff matrix suggested by Nowak and May \cite{nowak_n92b} where the reward of mutual cooperation is $R=1$ while mutual defection yields $P=0$ income. A cooperator gains $S=0$ payoff when facing a defector while a successful defector gets the temptation to defect ($T=b$). We investigate an extended parameter space $0<b<2$ to explore the SH region, too, while $S=0$ refers to the so-called weak PD for $1<b<2$.

For the MC simulations, we use random initial strategy distribution where both $C$ and $D$ strategies are present with the same frequency. Beside the strategy, each player possesses another parameter describing his willingness to make rational decision: having individual adoption rule ($K_x$). This parameter can be interpreted as a personal noise parameter as it contains the possible uncertainty factors in the strategy adoption. Such factors can emerge from the fluctuation of payoff parameters, changing of environment, errors in decision, individual freedom to risk a given amount of income, etc. depending on the situation which is modelled.
In our model, initially, we associate an adoption parameter $K_x$ to every player from a finite set, that is, $K_x \in \{K_1, K_2, \ldots , K_n\}$ where $n$ denotes the number of different $K$ values.

During the evolutionary process, we choose two neighboring players ($x$ and $y$) randomly, and we calculate their accumulated payoffs ($P_x$ and $P_y$) gained from PD games with their neighbors. In an elementary evolutionary step, player $x$ can adopt the strategy 
$s_y$ and/or the noise value $K_y$ of player $y$ with the probability
\begin{equation}
W =\frac{1}{1+\exp[(P_x-P_y)/K_x]}\,\,. \label{eq:prob}
\end{equation}
The possible adoption of the strategy and the noise value
happens independently of each other, i.e., it is possible that only one of them is adopted in an elementary step. As the Darwinian principle dictates, for $P_y - P_x \gg K_x$, both the strategy and the adoption rule of player $y$ is very likely to be adopted. 

According to the proposed protocol,
the adoption rule of player $y$ ($K_y$) can still be adopted even if the strategies are the same ($s_x = s_y$). As a consequence, the adoptions of strategies and rules can end independently arriving to one of the absorbing states formed by identical strategies and/or uniform adoption rules. In the absence of mutation the system cannot leave these states.

In general, the existence of many absorbing states can cause technical difficulties in the interpretation of numerical results achieved on small systems. For small sizes the system evolve quickly into one of the absorbing states despite the existence ("long-time stability") of a mixed state in the limit $L \to \infty$. These difficulties can be avoided by using sufficiently large system sizes that increases the duration of simulations. We will show, however, that the significantly faster simulations on small systems can also be utilized to extract accurate quantities characterizing the behavior of the present evolutionary games.

As we stressed, we used the same adoption probability for both evolving quantities determined by Eq.~\ref{eq:prob}. However, the time scales of evolutions can be distinguished, namely the strategy or the noise parameter may evolve faster or slower. Such a time scale separation of coevolving quantities was already studied in several earlier works \cite{zimmermann_pre04,vansegbroeck_prl09}. An interesting observation was the shift of effective payoff elements if the link of players, as an evolving quantity, change much faster than their strategies. In our case the time separation can be easily done by adding a multiplicative $0 < Q < 1$ prefactor to the transition rate of evolving quantity resulting a slower evolution comparing to the other one. Simulations show the fixed $K^\star$ noise value is robust: the system arrives into the same state independently on the time scale separation of evolving quantities. The fixation time, however, depends strongly on the applied $Q$ parameter. Accordingly, the results presented in the next sections are corresponding to the $Q=1$ case.

Starting the system from a random distribution of strategies and noise parameters the iteration of the above elementary processes governs an evolutionary process that can be quantified by recording the fraction $\rho_C(t)$ of cooperators as well as the portion $\nu_{K_i}(t)$ of players following the dynamical rule $K_i$. In one time unit (called MC step, in short, MCS) each player has a chance once on average to adopt a strategy and/or noise value of a neighbor.

Finally we mention that for homogeneous dynamical rule ($K_x = K$,  $\forall x$) the present model is
identical to a previously studied spatial model
\cite{szabo_pre05} as discussed briefly below.
Furthermore, for homogeneous strategy distribution [$s_x = C$ (or $D$),  $\forall x$] all the players receive the same payoff [$P_x=4$ (or $0$), $\forall x$] and the adoption probability (\ref{eq:prob}) becomes uniform, i.e., $W=1/2$ and the evolution of the $K_x$ distribution becomes identical to the process described by the $n$-state voter model \cite{liggett_85}.

\section{Selection of noise level on square lattice}

\begin{figure}[ht]
\centerline{\epsfig{file=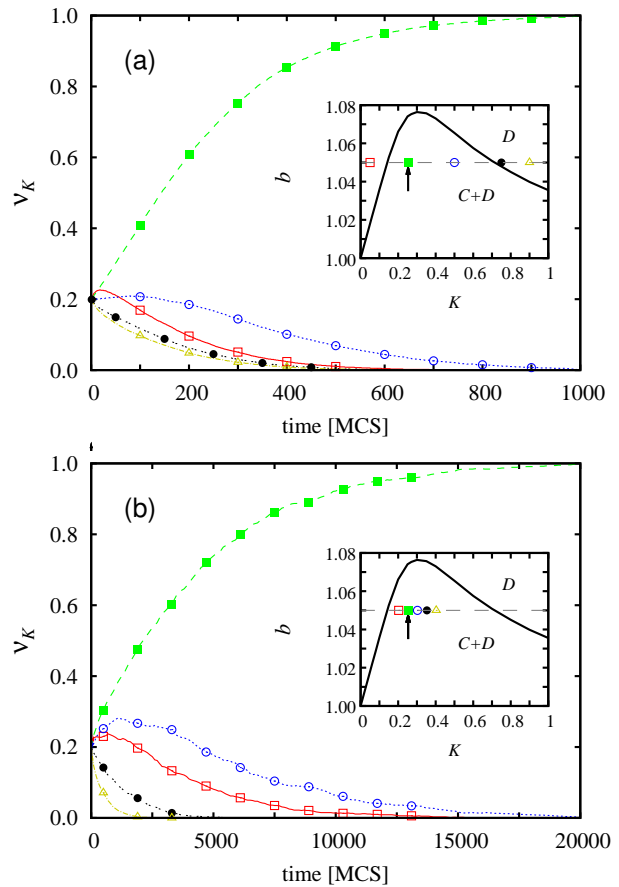,width=8cm}}
\caption{\label{selection}(color online) The time dependence of
the fraction of initial $K_i$ values ($\nu_K$) demonstrate the
selection of the distinguished adoption rule $K^{\star}$
(indicated by arrow in the inset) if four other rules ($K_i$) 
are permitted in the initial state on the square lattice at $b=1.05$
(upper panel). The inset, as part of the $b-K$ phase diagram,
shows the initial $K$ values chosen from the coexistence ($C+D$) and
absorbing ($D$) region as well. Lower panel illustrates the same
evolution if the additional four dynamical rules are closer to
$K^{\star}$, namely, $K_i= K^{\star}+0.05 (i-2)$ for
$i=1,...,n=5$. For both cases the linear system size was $L =
1000$.}
\end{figure}

\begin{figure}[ht]
\centerline{\epsfig{file=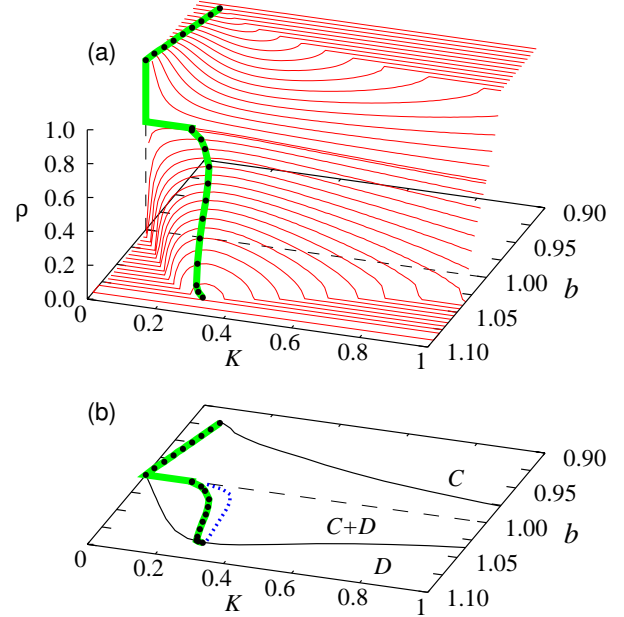,width=8cm}}
\caption{\label{sqr_mc}(color online) The surface of stationary cooperator density ($\rho$) as a function of $b$ and $K$ parameters when homogeneous $K$ distribution is supposed on square lattice. Closed circles and thick (green) line show the position of $K^{\star}(b)$ favored by the Darwinian selection of dynamical rules for fixed $b$. In the lower panel $K^{\star}(b)$ and the phase boundaries (separating the homogeneous $D$, the mixed $C+D$, and the homogeneous $C$ phases) are projected on the $b-K$ plane. For comparison, the positions of local maxima in the cooperator density $\rho$ for fixed $b$ are denoted by the dotted (blue) line. The dashed (black) line is just to mark the border between weak PD and SH games at $b=1$.}
\end{figure}

First we briefly outline the outcome if all players have the same adoption parameter ($K_x=K$,  $\forall x$), i.e., if only the strategy distribution can evolve \cite{szabo_pre05,vukov_pre06,vukov_pre08}. In this case, on lattices as underlying graphs, we can usually distinguish three regions when increasing $b$ for a fixed $K$ parameter.
Only cooperators remain in the final state after a transient time if $b<b_{c1}(K)$.
On the contrary, for $b_{c2}(K)<b$, defectors prevail. While the $C$ and $D$ strategies coexist in the region $b_{c1}(K)<b<b_{c2}(K)$ where the concentration of cooperators decreases continuously from 1 to 0 when increasing $b$ from $b_{c1}(K)$ to $b_{c2}(K)$. The continuous phase transitions at the two threshold values belong to the directed percolation universality class \cite{szabo_pre98,chiappin_pre99}. For uniform $K$ values the major feature of this system can be summarized in a $b-K$ phase diagram \cite{szabo_pre05} where the curves $b_{c1}(K)$ and $b_{c2}(K)$ denote the phase boundaries separating the homogeneous $D$, the coexisting $C+D$, and the homogeneous $C$ phases as it is partly illustrated in the inset of Fig.~\ref{selection} or Fig.~\ref{sqr_mc}b.

The systematic analysis of the proposed evolutionary process has justified the existence of a distinguished noise level $K^{\star}(b)$ within the $(C+D)$ coexistence region for any fixed $b$ (on square lattice the corresponding region is $b_{\rm min}=0.940(3) < b < b_{\rm max}=1.078(1)$ where the borders are the minimum and maximum values of $b_{c1}(K)$ and $b_{c2}(K)$ functions.). 
Figure~\ref{selection}a 
shows what happens if $K^{\star} \in \{K_1, K_2, \ldots , K_n\}$ for $n=5$. In this case the final state with players using the same $K^{\star}$ rule is reached after about 1000 MCS independent of $L$, if $L \agt 500$.

The speed of relaxation towards the homogeneous $K_x=K^{\star}$ state is strongly influenced by the initial set of possible $K_i$ values. The lower plot of Fig.~\ref{selection} illustrates a situation where the four additional $K_i$ values are very close to the distinguished $K^{\star}$ value.
Although the evolution of adoption rules is still straightforward, but
the relevant increase in the relaxation time may be related to the smaller difference in driving force favoring $K_x=K^{\star}$ at the expense of others. Similar slowing down can be observed if $n$ is increased while the maximal value of $K_i$ is limited. Anyway, the upper limitation of the range of possible $K_i$ values [(assuming that $\max (K_i) > K^{\star}$] does not influence the final results because the 
strategy adoption using 
the largest $K_i$ value dies out first (see Fig.~\ref{selection}).

\begin{figure}[ht]
\centerline{\epsfig{file=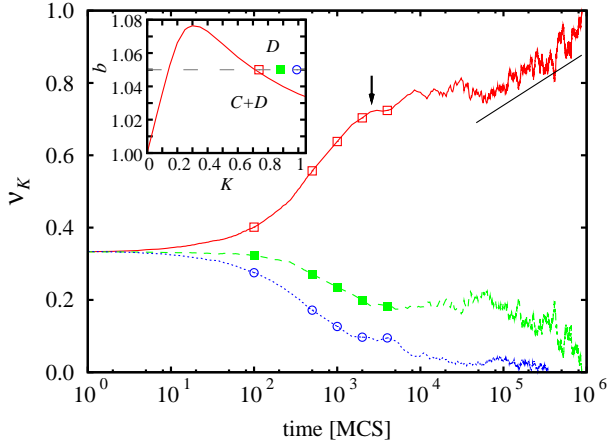,width=8cm}}
\caption{\label{voter}(color online) Time dependence of the fraction of the three $K_i$ values on square lattice at $b=1.05$ and $L=1000$. The inset shows the position of the initial $K_i$ values in the homogeneous $D$ region. The arrow points to the time $t_{\rm ext}$ when the $C$ strategy became extinct.
Solid line shows a $\log t$ function illustrating logarithmic coarsening in the final period.}
\end{figure}

The above results raise the question: What happens if the initial set of dynamical rules is out of the range of $(C+D)$ coexistence phase? (It also happens if $b < b_{\rm min}$ or $b > b_{\rm max}$.)
In this case the cooperators (or defectors) die out soon and the adoption of noise levels becomes random as it is described by the voter model predicting a behavior dependent on the spatial dimension $d$. Namely, algebraically growing domains of the same $K_i$ values occur on the one-dimensional lattice ($d=1$), the typical size of homogeneous domains increases with $\ln (t)$ if $d=2$, and the system remains inhomogeneous for $d \ge 3$
\cite{liggett_85,dornic_prl01}.
According to the above theoretical prediction a very slow (logarithmic) coarsening will be observed. This case is demonstrated in Fig.~\ref{voter}, where the initial $K_i$ values are in the all $D$ phase.
For the given simulation the cooperators have died out at $t=t_{\rm ext}=2600$ MCS resulting the conditions of voter model. Afterwards a coarsening without surface tension starts represented by huge fluctuations in $\nu_{K_i}$ functions. The semi-log scale demonstrates clearly that the absorbing state is reached after a long coarsening time. (To drive the eye we have plotted a $\log t$ function, too.) In the final state all the players use the same noise level. In principle, any of the persistent rules ($K_i$) can invade the whole finite system with a probability $\nu_{K_i}(t_{\rm ext})$ due to the fluctuations.

Several simulations were performed to study the cases where the initial $K_i$ values are positioned in both sides of the range of $(C+D)$ coexistence. The results indicated a qualitatively similar behavior plotted in Fig.~\ref{voter}.

\begin{figure}[ht]
\centerline{\epsfig{file=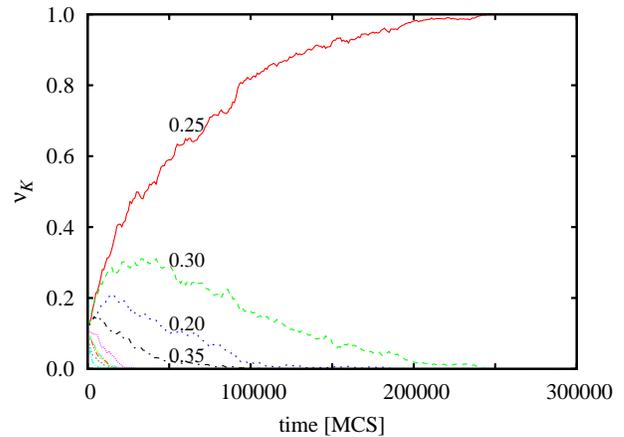,width=8cm}}
\caption{\label{mutation}(color online) Time dependence of the fraction of ten $K_i$ values, distributed equidistantly from $K_1=0.05$ to $K_{10}=0.5$, on square lattice in the presence of rare mutations ($\varepsilon=0.02$) at $b=1.1$ and $L=1000$. Labels indicate the closest $K_i$ values to $K^{\star}$.}
\end{figure}

Beside the large fluctuations for $t> t_{\rm ext}$ Fig.~\ref{voter} shows a smooth and deterministic variation in $\nu_{K_i}(t)$ within the period where both $C$ and $D$ strategies exist. This feature has inspired us to study the effect of strategy mutation on the Darwinian selection of noise levels in the regions where only cooperators ($b < b_{\rm min}$) or defectors ($b > b_{\rm max}$) would remain alive if the evolution was controlled by only imitations. For this purpose, in a few simulations, the above mentioned evolutionary rule is extended by allowing each player to change her strategy (from $C$ to $D$ or conversely) with a small probability $\varepsilon$. Figure~\ref{mutation} shows that slightly above $b_{\rm max}$ the Darwinian selection favors also a distinguished rule $K^{\star}(b=1.1) \simeq 0.25$ (in the presence of rare mutations) that can be considered as the analytical continuation of $K^{\star}(b)$ obtained within the coexistence region. Evidently, the favored rule depends on both $b$ and $\varepsilon$. As the spreading of the distinguished noise level is catalyzed by the mutants, the speed of this process vanishes with $\varepsilon$.

If we take $b < b_{\rm min}$ value, being in the SH region, the introduction of small mutations will result a fixation value that can be also considered as an analytical continuation of fixed $K^{\star}$ values from the $b > b_{\rm min}$ interval. Namely, only small $K \approx 0$ values survive and the fixation time increases drastically as $b$ decreases.
Instead of the further analysis of the effect of mutations henceforth our attention will be focused on the
behavior of $K^{\star}(b)$ in the $(C+D)$ coexistence phase in the absence of mutation.

The above mentioned features of the selection of dynamical rules refer to serious technical difficulties (related to the long runs on large lattice) in the determination of $K^{\star}(b)$ with an adequate accuracy. It turned out, however, that this quantity can be evaluated more efficiently by repeating simulations with only two possible $K_i$ values on small systems. In this case ($n=2$) we choose a simple notation, namely $K_1 = K - \Delta K /2$ and $K_2 = K + \Delta K / 2$. For small sizes the random initial state evolves rapidly into one of the absorbing phases where all the players use uniformly the 
value $K_1$ or $K_2$. Starting from different (random) initial states these simulations are repeated many times (typically $N_r=2000$) and the preference of the second rule ($K_2$) is measured by the quantity $f=g(K_2)-g(K_1)$ where $g(K_i)$ is the probability that the evolution ends up in the absorbing state with players using uniformly the rule $K_i$. As the simulations hold until reaching one of the absorbing states therefore $g(K_1)+g(K_2)=1$ and $f$ varies from -1 to +1. Similar quantities are used frequently for the investigations of finite systems within the framework of Moran process \cite{moran_62}.

\begin{figure}[ht]
\centerline{\epsfig{file=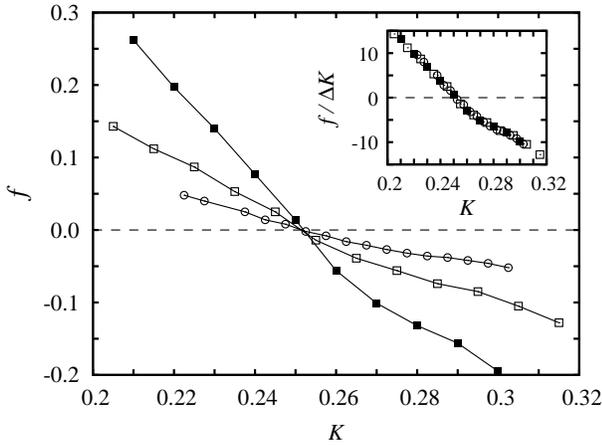,width=8cm}}
\caption{\label{drive}MC results for the measure $f$ of preference between two dynamical rules as a function of $K$ on a square lattice of $40 \times 40$ sites for $b=1.05$ if $\Delta K=0.005$ (circles), $\Delta K=0.01$ (open boxes), and $\Delta K=0.02$ (closed boxes).
Inset: data collapse when scaling $f$ by $1/ \Delta K$ suggesting the identification of the preferred adoption rule is independent of the magnitude of $\Delta K$.}
\end{figure}

Figure \ref{drive} demonstrates the results of these investigations when varying the value of $K$ and $\Delta K$ for a fixed system size $L$ and $b$.
According to MC data the position of $K^{\star}$, where the sign of $f$ changes, is independent of the value of $\Delta K$ within the statistical error. $f$ varies linearly with $K$ in the close vicinity of $K=K^{\star}$ [more precisely, $f \sim (K^{\star}-K$)]. Naturally, larger $\Delta K$ involves larger difference in the ``driving force'' favoring one of them. This fact can be demonstrated by the data collapse when choosing a more suitable scale for the vertical axes.

\begin{figure}[ht]
\centerline{\epsfig{file=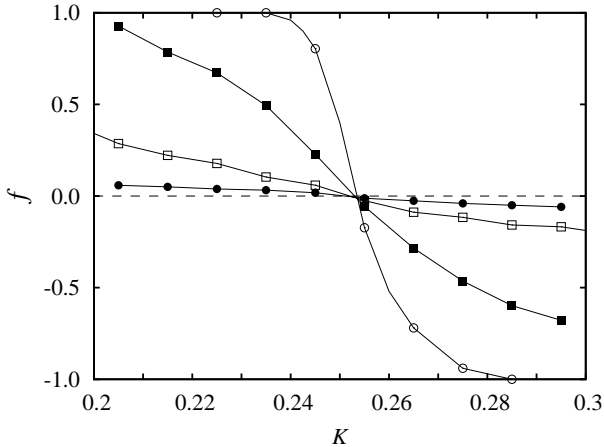,width=8cm}}
\caption{\label{fsa_sqr} The measure of preference between the dynamical rules of $K_1$ and $K_2$ versus $K$ on a square lattice for different sizes [$L= 160$ (open circles), $80$ (closed boxes), $40$ (open boxes), and $20$ (closed circles)] if $\Delta K=0.01$ and $b=1.05$.}
\end{figure}

Figure \ref{drive} shows that $f$ decreases smoothly
if $K$ is increased for $L=40$. Evidently, this transition becomes sharper if we choose larger systems as demonstrated in Fig.~\ref{fsa_sqr}. According to these data the system size influences only the absolute value of $f$ and the functions $f(K)$ becomes zero at $K=K^{\star}$ independently of the system size if it is large enough. In the limit $L \to \infty$ a step-like transition (from 1 to -1 at $K^{\star}$) is expected. To sum up, the position of $K^{\star}$ can be determined by using only one $\Delta K$ value at small system size, which reduces the necessary fixation time drastically.

The positive (negative) value of $f$ indicates the parameter region where the system can evolve towards larger (smaller) $K_x$ values through weak mutations in $K_x$ during the adoption processes. Within this context $K^{\star}$ in Fig.~\ref{drive} can be considered as an attractor.

The above features are utilized in the accurate determination of $K^{\star}$ for a given value of $b$. In our previous work \cite{szabo_epl09} these calculations are repeated to determine the distinguished rule ($K^{\star}$) for different $b$ values within the range of weak Prisoner's Dilemma. Now this analysis is extended to the whole range of $b$ (i.e., $b_{\rm min} < b < b_{\rm max}$) and the results are summarized in 
Fig.~\ref{sqr_mc}b.

In order to get deeper understanding about the fixation of noise levels, the
cooperator density [$\rho(K)$] profiles are illustrated above the
whole $b-K$ plane (see Fig.~\ref{sqr_mc}a). The
plotted $\rho(K)$ curves are obtained for fixed $b$. In the
simulations we used the following system parameters: system sizes
of $10^5$-$10^6$ players, relaxation time of $5 \cdot 10^5$-$10^6$
MCS and $10^5$-$2 \cdot 10^5$ more MCS for averaging to get the
steady state cooperator density values. Larger system sizes and
longer simulation times were needed in the vicinity of the
critical points. The statistical errors of the plotted data are
comparable with the line thickness.

Each plotted $\rho(K)$ curve (for fixed $b$) shows a local maximum in the range of weak PD ($b>1$). 
The plot of Fig.~\ref{sqr_mc}b demonstrates clearly that the values of $K^{\star}(b)$ are close to the site where the cooperator density $\rho$ has a local maximum for the given $b$. Although the position of the local maxima in the average payoff and $\rho$ (for fixed $b$) are distinguishable, the difference between these positions is small and comparable with the size of symbols as discussed in \cite{szabo_epl09}.

Within the region of SH game ($b<1$) Fig.~\ref{sqr_mc} shows a significantly different behavior. First we emphasize that the surface $\rho(K,b)$ exhibits a valley within the coexistence region, otherwise $\rho =1$. More precisely, for fixed $b$ the curve $\rho(K)$ differs from 1 within the coexistence region [namely, if $K_{c1}(b) < K < K_{c2}(b)$ assuming that $b> b_{\rm min}$] and has a local minimum. It turned out that within this region the Darwinian selection of noise values
prefers also a distinguished rule that is positioned at the left edge of the "valley", that is, $K^{\star}(b) = K_{c1}(b)$. This means, that if we study a system with only two initial rules ($K_1$ and $K_2$ both within the coexistence region) then the smaller one will
spread in the whole system in the final state.

For uniform dynamical rule ($K_x=K$) on the square lattice with nearest neighbor interactions both $b_{c1}(K)$ and $b_{c2}(K)$ (the phase boundaries in 
Fig.~\ref{sqr_mc}b) go to 1 if $K$ tends to either 0 or $\infty$ in such a way that one can observe an optimal noise level for the cooperators in the PD region and another one for the defectors in the SH games. 
In other words $\rho(K)$ curves have a local maximum (minimum) in the PD (SH) region. As it was shown by a previous study \cite{vukov_pre06} the local maximum of cooperation level
is related to the absence of overlapping triangles (three-site cliques) 
of interaction graph.
The comparison of the mentioned surface and the fixation values of $K_i$ for both games in 
Fig.~\ref{sqr_mc}
suggest that the possible evolution of strategy adoption rule will drive the system into a state that ensures closely the optimal cooperation level independently of the studied dilemma game.
In the subsequent section we will consider another type of connectivity structure exhibiting a slightly different behavior.

\section{Results on kagome lattice}

In real human connectivity structures a relevant portion of the neighbors of a player $x$ is also connected to each other (that is, the so-called clustering coefficient is sufficiently large) \cite{watts_n98}. The main effect of this topological feature can be well investigated if the players are distributed on the sites of the two-dimensional kagome lattice where each player has also four neighbors.
The latter feature makes possible to exclude the additional impact by changing coordination number.
The systematic investigation of the evolutionary PD game on this connectivity structure has explored a basically different phase diagram in comparison with that observed on the square lattice \cite{szabo_pre05}. It turned out that the upper threshold value of temptation [$b_{c2}(K)$] decreases monotonously from $3/2$ to 1 if $K$ is increased from 0 to $\infty$. Due to
the different limit
of $b_{c2}(K)$ threshold values, one can also expect basically different behavior in the Darwinian selection of
noise levels for the PD.

The monotonous $K$ dependence of $\rho$ 
in the low noise limit is related to presence of the overlapping triangles that support the spreading of cooperators through the lattice \cite{vukov_pre06}. In fact, the MC simulations indicate similar behavior for many other regular connectivity structures (including two- and three-dimensional lattices and some other regular networks) where the overlapping triangles span the whole system. Thus the kagome lattice can be considered as a sample representing the latter type of 
interaction graphs.

Here it is worth mentioning that the low $K_i$ (or $K$) values lead to diverging simulation times and cause technical difficulties in the quantification of the low noise behavior. The subsequent results were extracted from a set of simulations where $K_i \ge K_{min}=0.002$ is chosen for all $i$.

\begin{figure}[ht]
\centerline{\epsfig{file=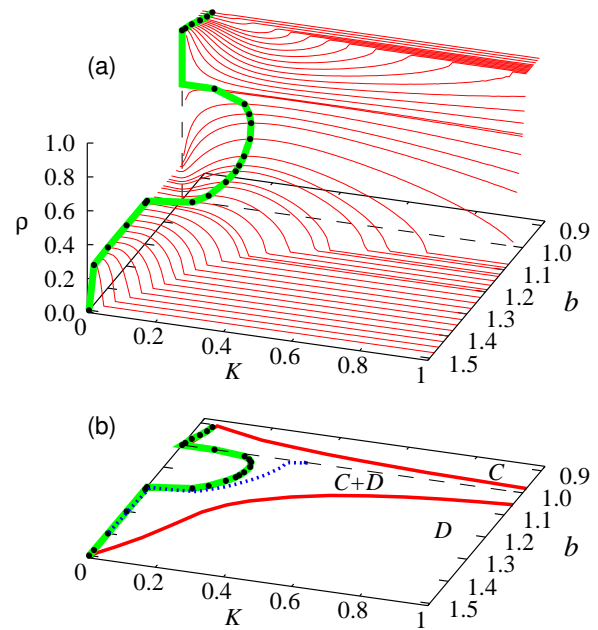,width=8cm}}
\caption{\label{kag_mc} The same plot as Fig.~\ref{sqr_mc} for kagome lattice. The cooperation level (upper panel) has minima in the SH region and maxima in the PD area. The maximum at positive $K$ decreases as $b$ increases and is replaced by a maximum at $K=0$ if $b$ exceeds a threshold value ($b_{th}=1.182$). Bullets and thick green line mark the main attractor while the positions of maximal cooperation level are also denoted by dotted blue line.}
\end{figure}

As we observed in case of square lattice topology, the fixation of evolving $K$ values is in close connection to the surface of maximal cooperation when using homogeneous dynamical rules ($K_x=K$). Therefore, the same surface on $K-b$ plane is determined for kagome lattice as plotted in Fig.~\ref{kag_mc}a.

For the PD game ($b>1$) the $\rho(K)$ curves (for fixed $b$) vary continuously from nonzero value of $\rho(K=0)$ until reaching the absorbing state ($\rho=0$). We can distinguish different types of behaviors although the accurate separation of the corresponding regions of parameter is prevented by the above mentioned technical difficulties. From the extrapolation of the low noise behavior monotonously decreasing $\rho(K)$ can be concluded if $K$ is increased from zero to infinity at $3/2>b \agt 1.4$. In the subsequent region, $1.4 \agt b > b_{th}=1.182$, the curves $\rho(K)$ possess only one local maximum close to $K=0$. At $b=b_{th}$, with a sudden jump there appears another local maximum while the other local maximum close to $K=0$ still exist to $b=1$. The absolute maximum is the one belonging to larger $K$ value. This behavior indicates that the triangle percolation added a support for cooperation for lower noise values and the cooperator density profile can be derived as the superposition of a plateau originated from the triangle percolation effect and the normal one-peak profile of a lattice. In the region of SH game ($b<1$) the $\rho(K)$ curves are resembling those discussed in the previous section. As well as for the square lattice the function $b_{c1}(K)$ goes to 1 if $K$ tends to zero or infinity.

\begin{figure}
\centerline{\epsfig{file=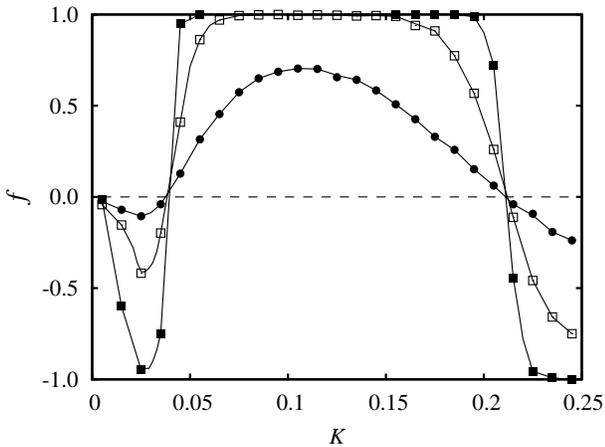,width=8cm}}
\caption{\label{fsa_kag} The same plot as Fig.~\ref{fsa_sqr} but using kagome lattice topology at $b=1.1$ and $\Delta K=0.01$. The system sizes are $3 \times L \times L= 3 \times 20 \times 20$ (closed circles), $3 \times 40 \times 40$ (open boxes), and $3 \times 80 \times 80$ (closed boxes). The plot demonstrates clearly the existence of two attracting fixed points at $K^{\star}=0$ and at $K^{\star}=0.212$ where the border of attraction of fixation values is at $K_{sep}=0.038$.}
\end{figure}

Our conjecture, based on the close relation of the fixed noise level and the optimal cooperation level of homogeneous $K$ system, is completely supported by the evolution of adoption rules on kagome topology, too. More precisely, in the SH region the $K_i$ values of coexistence $C+D$ phase drift to the minimal $K$ value to reach $K_{c1}(b)$ that ensures the maximal cooperation ($\rho = 1$). Technically, if we choose two initial $K_i$ values from the coexistence region than the lower $K$ value will spread eventually in the whole population. At high values of $b$ in the PD region the final value of noise parameter is always the lowest among the initial set signalling $K^{\star} \approx 0$. This feature is related to the fact that cooperation level always has a local maximum at $K \approx 0$. Decreasing $b$, however, a bifurcation occurs at $b=1.185$: besides the $K^{\star}\approx0$ fixed point a new attractor appears located at a positive $K$ value. The presence of two attractors are demonstrated in Fig.~\ref{fsa_kag} where the measure of preference, is plotted by means of $K$. In this $b$ region the initial $K_i$ values always destine at $K^{\star}\approx 0$ if they are below $K_{sep}(b)$ value. (The position of separator, as the border of basins of attractors will be discussed in the next section.) If $K_i > K_{sep}$, the adoption rules converge to the above mentioned $K^{\star} \neq 0$ value. Naturally, if all $K_i > K_{c}$, means all $K_i$ values are from the absorbing $D$ phase, similar behavior can be observed as shown in Fig.~\ref{voter}.

An interesting situation occurs when the initial $K_i$ values are distributed from the whole $(0, K_{c})$ interval. In this case, only $K^{\star} \neq 0$ fixed point survives, signalling that the latter is the stronger attractor. As expected, the position of positive $K^{\star}$ attractor is close to the $K$ value where maximal cooperation level is measured for homogeneous $K$ model. Summing up our observations for both representative topologies and for both dilemma games, it is concluded that the system spontaneously will evolve to a state that is favorable for cooperation if the adjustment of noise (strategy adoption) is allowed.

\section{Dynamical cluster approximations}

Beside the MC simulations, we performed dynamical cluster approximations \cite{szabo_pr07} on kagome lattice. The choice of kagome lattice for this type of investigation was motivated by its simplicity. To highlight the difficulties in the application of these sophisticated technique first we emphasize that neither the mean-field (one-site) nor the pair (two-site) approximations were capable to give an adequate 
description of the homogeneous $K$ model, particularly in the low noise limit. On the square lattice,
a higher level of approximation (four-site approach) is needed to reproduce qualitatively the results of MC simulations.
Furthermore, a more demanding nine-site level is necessary to reach an adequate accuracy. On the contrary, the three-site (triangular) approximation can reproduce quantitatively well the results of MC simulations on kagome lattice.

Now we briefly survey this method for a simple case where each site $x$ has only two states ($s_x=C$ or $D$) and later we give the main details of the extension to the four-state systems
that is necessary to describe our present model
(for a more detailed description see the papers \cite{szabo_pr07,dickman_pre01,szabo_pra91} with further references therein). Within the framework of this approximation the system is characterized by all possible configuration probabilities $p_3(s_{\alpha},s_{\beta},s_{\gamma})$ on a block of three neighboring sites forming a regular triangle on the kagome lattice ($\alpha$, $\beta$ and $\gamma$ are site labels within the three-site block). In the present case the eight possible configurations can be given using by only three parameters due to the symmetries and compatibility conditions. These configuration probabilities are determined by solving a set of differential equations expressing the derivative of $p_3(s_{\alpha},s_{\beta},s_{\gamma})$ with respect to time (denoted as $\dot{p}_3(s_{\alpha},s_{\beta},s_{\gamma})$). The main difficulties in the application of this method comes from the fact that the contribution of the elementary processes (here strategy adoption between two neighboring sites) to the quantity $\dot{p}_3(s_{\alpha},s_{\beta},s_{\gamma})$ depend on the configuration containing seven sites.

\begin{figure}[ht]
\centerline{\epsfig{file=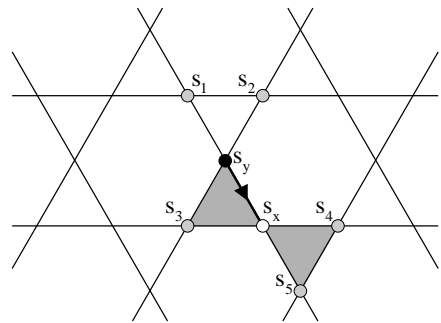,width=6cm}}
\caption{\label{kag_dmf} Strategy adoption from $s_y$ to $s_x$ (marked by an arrow) will modify the three-site configuration probabilities on triangles marked by grey color. All the other neighbors influencing the payoff difference are denoted by grey circles.}
\end{figure}

Figure \ref{kag_dmf} illustrates an elementary process with the neighborhood affecting the probability of strategy adoption from site $y$ to $x$. This process decreases $\dot{p}_3(s_{x},s_{y},s_{3})$ and simultaneously increases $\dot{p}_3(s_{y},s_{y},s_{3})$ with a value
\begin{equation}
\Delta p =W p_7(s_1, \ldots , s_5)
\label{eq:deltap}
\end{equation}
where $W$ describes the payoff dependence defined by (\ref{eq:prob}) and $p_7(s_1, \ldots , s_5)$ denotes the seven-site configuration probability. For the the present connectivity structure the latter quantity can be approximated by a Bayesian formula:
\begin{equation}
p_7(s_1, \ldots , s_5) \simeq { p_3(s_1,s_2,s_y)p_3(s_y,s_x,s_3)p_3(s_x,s_4,s_5) \over
p_1(s_x) p_1(s_y)}
\label{eq:p7}
\end{equation}
where the one-site configuration probability in the denominator can be also expressed by the three-site configuration probabilities as
\begin{equation}
p_1(s_{\alpha}) = \sum_{s_{\beta},s_{\gamma}}  p_3(s_{\alpha},s_{\beta},s_{\gamma}) \, .
\label{eq:p1}
\end{equation}
Summarizing the contribution of all the possible elementary processes affecting the values of $\dot{p}_3(s_{\alpha},s_{\beta},s_{\gamma})$ one can derive a set of differential equations. Now we do not wish to display the huge formulae depending only on the three-site configuration probabilities due to approximative formula (\ref{eq:p7}). Instead of it we emphasize that one can easily develop a computer algorithm to collect systematically all the contributions and the resultant formulae can be used to find the stationary solution(s) numerically for any values of parameters. In the knowledge of the stationary solutions of the three-site configuration probabilities the most relevant characteristic of the stationary state can be evaluated, for example, $\rho = p_1(C)$. Using this approach the $b-K$ phase diagram was reproduced qualitatively well in \cite{szabo_pre05,vukov_pre06}.

In the present work this method is extended by substituting $(s_{\alpha}, K_{\alpha})$ for $s_{\alpha}$ where $s_{\alpha}=C$ or $D$ and $K_{\alpha}=K_1=K-\Delta K /2$ or $K_2 = K+ \Delta K /2$. This extension does not influence the applicability of the above described method.  Using only two initial adoption rules, it was possible to keep the number of the feasible configurations low enough so that the numerical solution of the differential equation system was fairly fast. It turned out that this method is capable to reproduce all the relevant features characterizing the Darwinian selection of the dynamical rules. For the quantitative analysis we used small $\Delta K = 0.001$ values to evaluate the position of the attractor ($K^{\star}$) and the separatrix for any values of $b$.

\begin{figure}[ht]
\centerline{\epsfig{file=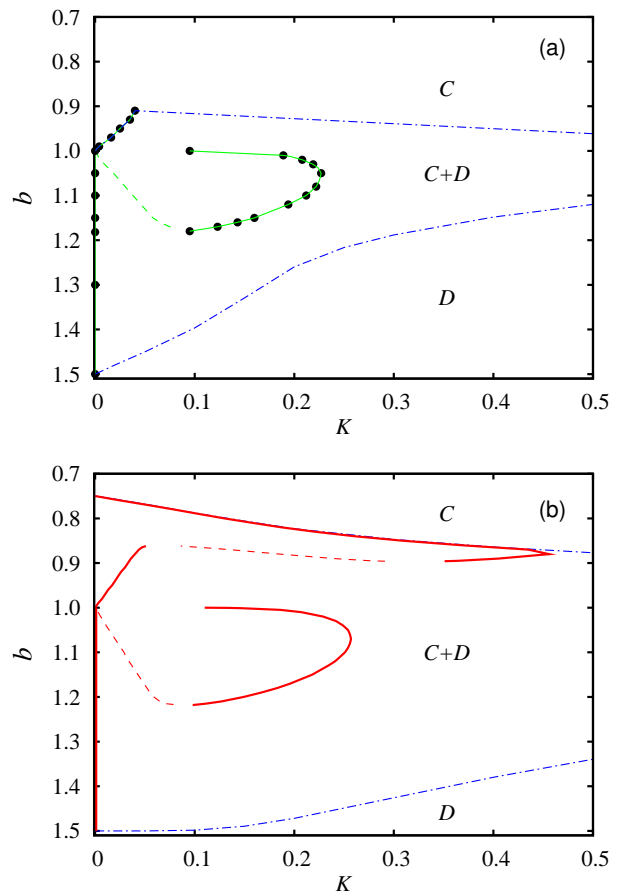,width=8cm}}
\caption{\label{kag_3p}(color online) Positions of fixed points of adoption rules on $b-K$ plane as predicted by MC simulation (top) and three-site cluster approximation (bottom) for kagome lattice. Bullets, connected by solid (green) line, show the fixed point given by MC simulations. The position of separatrix in the two-attractor region is denoted by dashed (green) line. The borders of phases are marked by dashed-dotted (blue) lines. In the lower panel solid (red) lines show the position of attractors while separatrix is denoted by dashed (red) lines. Dashed-dotted (blue) lines mark the borders of phases.}
\end{figure}

The results are summarized and compared with the MC data in Fig.~\ref{kag_3p}. The predictions of the generalized mean-field approximation for the PD region are in excellent agreement with MC data. As both approaches stated, there are two attractors in a restricted interval of $b$ where a separatrix marking the border of basins of attractors tends to $b=1$ if $K$ decreases. It also means that the basin of attractor $K^{\star}\approx 0$ keeps getting wider with increasing $b$. When separatrix reaches the other fixed point the latter disappears resulting a unique fixed point in the large $b$ region. Turning to the SH side, the dynamical cluster approximation predicts increasing $K^{\star}$ fixed point as we leave $b=1$, which is again in nice agreement with the MC results. Further decreasing $b$, however, the approximation predicts a small two-attractor region and finally the $K^{\star}(b)$ function coincides with the phase boundary where $\rho$ becomes 1. 

Before evaluating the predictions of the present cluster mean-field approach, we should stress that the simpler (two-state) version of the three-site approximation, which is valid for homogeneous $K$ case, cannot describe correctly the functions $\rho(K)$ as well as $b_{c1}(K)$ in the low $K$ limit. Namely, the border of all C phase tends to $b=0.75$ instead of $b=1$ when $K \to 0$. (According to this, defectors can survive the deterministic limit when $b<1$.) Such a qualitative failure of the approximation might have been the consequence of the small number of independent variables. In other words, the restricted freedom prevents the approximation to find the valid solution. At the same time, if we increase the number of independent variables by letting different $K_i$ values for players, the approximation is already capable to find the relevant solution (at least in the vicinity of $b=1$). Our argument is supported by the fact that the five-site approximation for the homogeneous $K$ model can also describe the behavior $b_{c1} \to 1$ in the $K \to 0$ limit. The relevant increase in $b_{c1}$ can remove the artifact(s) occurred at low $b$ values. Summing up, the extension of the three-site approximation by letting different $K_i$ values was capable to indicate the correct results of the more sophisticated approximation based on larger cluster of sites.

\section{Discussion and outlook}

We studied evolutionary Prisoner's Dilemma and Stag Hunt games on two 
representative two-dimensional lattices. The underlying structures were the square lattice and the kagome lattice exemplifying spatial connectivity structures without or with triangle percolation. We analyzed the 
simultaneous evolutions of strategy and a player specific noise parameter
used by individuals in the strategy imitation processes. It turned out that players prefer to use the same distinguished noise level, namely the same way of strategy adoption
and the fixed points of the adoption parameter are systematically close to those parameter values which are the most favorable for cooperation for the PD games. It implies that the evolution of the adoption parameter drives the population to a state which assures substantial cooperation within the coexistence region. In the region of SH game the maximum average payoff is achieved at the peripheries of the coexistence region. It is shown that in the latter case the Darwinian selection favors the edge of coexistence region where the noise level ($K$) is lower. The preference of the lower noise level can also be observed within the PD region because the distinguished rule $K^{\star}$ was always smaller than that where the maximum average payoff (or cooperator density) occurs for homogeneous dynamical rules at fixed payoff. The above results raise many general questions about the main features of states (including many aspects of the model itself) favored by the Darwinian selection. Here we emphasize that the Darwinian selection seems to be more efficient within the coexistence region where the simultaneous evolution of strategies accelerates the evolution of adoption rules, too.
The latter observation is confirmed by simulations where the coexistence in maintained artificially by introducing rare mutations in the systems.

The above investigations have required to improve the accuracy particularly in the low noise limit where the relaxation time diverges. The systematic investigation of the cooperator's density versus $b$ (temptation to choose defection) and $K$ have indicated different types of non-analytical behaviors in the limit $b \to 1$ and $K \to 0$ as Figs.~\ref{sqr_mc} and \ref{kag_mc} show. 
At the same time we found qualitatively similar behavior on both structures in the region of SH game. Although most of these features can be reproduced qualitatively well by the dynamical cluster methods on sufficiently large cluster of sites, we think that further analysis is required to clarify the effects of the sucker's payoff $S$, topology of connectivity network, and dynamics (e.g., when irrational choices are forbidden) on the non-analytical behavior appeared here at $b=1$ and $K=0$.

The present investigations have expanded the research of coevolutionary game theory by applying Darwinian selection among a continuous set of noise levels in dynamical rules used by the individuals while other relevant ingredients of the model were fixed.  In most of the previous studies of the coevolutionary games only two ingredients of the system were allowed to evolve simultaneously. Recently, Van Segbroeck {\it et al.} have studied a model where the players could modify three quantities: their strategy, connections, and the way how a new partner is chosen \cite{vansegbroeck_prl09}. In the light of the latter model one can ask what happens if the payoff parameter $b$ is also considered as an individual property ($b_x$) and it can be adopted from the neighbors as well as the strategy $s_x$ and dynamical rule $K_x$. The preliminary MC results have indicated that within the strategy coexistence region of the $b-K$ plane the system evolves toward $K^{\star}(b)$ with favoring smaller $b$ values. It is found that the system evolves fast toward a state where players use game of the lower $b$ value and subsequently the homogenization in $K_x$ will be done as described above. This means that the Darwinian selection prefers SH to PD game, that is, the system develops an environment where the mutual cooperation (providing the maximum average payoff) can be achieved more conveniently. Similar results were reported by Worden and Levin \cite{worden_jtb07} and also by Fort \cite{fort_epl08} who studied models with different adoptions of payoff parameters. Evidently, other results can be obtained if 
the simultaneous evolution of the connectivity structure (interaction and learning networks \cite{ohtsuki_prl07}) is also possible.

In principle, all the ingredients of the multi-agent coevolutionary games can be the subject of Darwinian selection if we assume that these features are determined by the participants. In that case the system can evolve towards a strategy distribution with a proper connectivity structure, payoff parameter, adoption rule(s), mutation, {\it etc.}, that are preferred by the Darwinian selection. As a consequence the given Darwinian selection will show us the preferred features (or parameter values) we can fix when exploring the effect of other properties \cite{ho_jet07}. To be  more precise, the $K^{\star}$ value(s) can be suggested in numerical simulations if one wish to fix the noise level.

Finally we would like to mention that in many real systems besides the evolving individual's features there are external conditions affecting the system behavior. For example, the noise itself can arise from external sources as it was investigated by Traulsen \cite{traulsen_prl04} and Perc \cite{perc_njp06a}. Further systematic investigations are required to clarify the effect of the external noise or any other questions mentioned above.

\section*{Acknowledgments}

This work was supported by the Hungarian National Research Fund (Grant No. K-73449), Bolyai Research Grant and FCT Portugal.

\end{document}